\documentstyle[multicol,aps,psfig]{revtex}
\begin{document}

\preprint{\vbox{\hbox{DRAFT of U. of Iowa preprint 97-2504}}}

\title{Non-Universal Quantities from Dual RG Transformations}

\author{Y. Meurice and S. Niermann\\ 
{\it Department of Physics and Astronomy, University of Iowa, 
Iowa City, Iowa 52242, USA}}

\maketitle

\begin{abstract}
Using a simplified version of the 
RG transformation
of Dyson's hierarchical model, we show that one can
calculate all the non-universal quantities 
entering into the scaling laws by combining an expansion 
about the high-temperature fixed point with a dual expansion about
the critical point.
The magnetic susceptibility is expressed in terms of two dual
quantities transforming covariantly under an RG transformation
and has a smooth behavior in the high-temperature limit.
Using the analogy with hamiltonian mechanics, the simplified example
discussed here is similar to the anharmonic oscillator, while
more realistic examples can be thought of as coupled oscillators,
allowing resonance phenomena.
\end{abstract}
\begin{multicols}{2} \global\columnwidth20.5pc
\multicolsep = 8pt plus 4pt minus 3pt

One important contribution of the Renormalization Group (RG) method
is to show that 
there exists a close connection\cite{wilson} between statistical 
mechanics near criticality and Euclidean field theory
in the large cut-off
limit.
In this approach, the determination of the renormalized quantities
at zero momentum amounts to the determination of a certain number
of parameters appearing in the scaling laws. 
Some of these parameters are universal (the critical exponents) 
and much effort 
has been successfully devoted to their calculation.
On the other hand, new techniques need to be developed 
in order to reliably calculate the non-universal parameters. 

We limit here the discussion to the case of scalar 
field theories with a lattice regularization (spin models). 
This class of models has several important applications in particle 
physics (e.g., self-interactions in the Higgs sector) 
and condensed matter physics (e.g., ferromagnetism)
which require an accurate non-perturbative treatment.
For $\beta $, the inverse temperature (or the hopping parameter), close 
to its critical value $\beta_c$, 
one can express the magnetic 
susceptibility (zero-momentum two-point function) with
an expression which, when $D<4$, takes the form\cite{parisi}
\begin{equation}
\chi\simeq (\beta _c -\beta )^{-\gamma } (A_0 + A_1 (\beta _c -\beta)^{
\Delta }+\ldots )\ ,
\label{eq:param}
\end{equation}
where the non-universal quantities 
$A_0,\ A_1,\dots$ are functions of the other (``bare'') 
parameters of the theory.
Following the discussions of Ref.\cite{wilson,rapid}, 
we can use Eq. (\ref{eq:param})
to obtain a non-perturbative 
definition of the renormalized mass $m_R^2$ 
of the form
\begin{equation}
m_R^2={{\Lambda_R^2}\over{A_0+A_1({\Lambda_R\over
\Lambda})^{2\Delta \over{\gamma}}+\dots}} \ ,
\label{eq:cutoff}
\end{equation}
for a scale of reference $\Lambda_R$, and a UV cut-off $\Lambda$.
Similar considerations apply to the other renormalized quantities 
which can be obtained from the higher point functions. In order to 
complete in a quantitative way this non-perturbative renormalization
program, one needs to be able to calculate the non-universal quantities
in Eq. (\ref{eq:cutoff}) as well as the universal ones. 

This task can be achieved\cite{rapid}
in the case of the well-studied hierarchical model
\cite{hier,wittwer,osc,finite}. 
Using the numerical methods developed in Ref. \cite{finite},
one can calculate the susceptibility at various values of $\beta$
and extract 
the unknown parameters in Eq. (\ref{eq:param}) 
by direct fits\cite{rapid}. 
This is a rather tedious 
procedure involving successive numerical refinements.
A more satisfactory approach consists in expanding about the 
fixed point calculated by Koch and Wittwer 
\cite{wittwer}. In a system of coordinates where the fixed point
is at the origin and the axes coincide with the eigenvectors of the 
linearized transformation, the RG transformation reads
\begin{equation}
d_{n+1,m}=\lambda_md_{n,m}+\sum_{k,l}\Gamma_m^{kl}d_{n,k}d_{n,l}\ ,
\label{eq:quad}
\end{equation}
where the $\lambda_m$ are the eigenvalues of the linearized RG
transformation (which yield
the critical exponents) 
and the $\Gamma_m^{kl}$ are calculable coefficients.
In Ref. \cite{rapid}, we found that the direct fits and the linearization
agree with 12 significant digits for the leading 
exponent $\gamma$.
The linearization method does not provide a way to calculate
the non-universal quantities $(A_0,\  A_1\ldots)$. 
It would be a great accomplishment to show that 
these non-universal quantities could be calculated 
accurately
by taking into account 
the non-linear terms in Eq. (\ref{eq:quad}).
In this Letter we show 
that such a calculation can be performed in a one-variable version 
of Eq. (\ref{eq:quad}) which is justified in the next paragraph.
Furthermore, some of the calculations can be performed 
much more efficiently
by combining the
expansion described above with a dual expansion which can 
be identified with
the high-temperature expansion.

In the following, we consider 
a recursion relation for the magnetic susceptibility which reads
\begin{equation}
\chi_{n+1}=\chi_n +(\beta/ 4) ({c/2})^{n+1}\chi_n^2
\label{eq:sus}
\end{equation}
where $c=2^{1-2/D}$ in order to approximate a $D-$dimensional model
and $n$ stands for the fact that the susceptibility is 
calculated with a number of sites $2^n$.
In the following, we limit ourselves to a range of parameters 
corresponding to ferromagnetic interactions in the symmetric phase,
and such that an infinite volume limit exists. This means $0<\beta<\beta_c
$ (the value $\beta_c$ is calculated below)
and $0<c<2$.
Eq. (\ref{eq:sus}) 
can be obtained as follows. First, we
consider the recursion formula for the hierarchical model
in the approximation where the Fourier transform of the local
measure is approximated by a polynomial of degree 2 (this is
called $l_{max}=1$ in Ref.\cite{finite}) 
and then we expand the resulting recursion
formula for the susceptibility to first order in $\beta$.
The variable is then rescaled in order to obtain 
a recursion formula in terms of the physical quantity $\chi$.
The recursion formula Eq. (\ref{eq:sus}) becomes an accurate 
approximation of the exact  recursion formula for Dyson's model
when $n$ is large enough. A related formula is used in Ref. \cite{finite}
to estimate the finite volume effects (see Eq. (5.1) therein).
In the following, we use the notation $\xi$ for ${c/2}$.
For definiteness, we will take the initial value 
$\chi_{0} = 1$.

The explicit dependence on $n$ and $\beta$ in Eq. (\ref{eq:sus}) can be 
eliminated by introducing $
h_{n} \equiv \alpha \xi^{n} \chi_{n}$. The constant of proportionality $\alpha$
can be fixed by requiring that the fixed points of the the RG transformation 
in terms of the new variable are 0 and 1. This yields $\alpha=\beta
{c^2}/(8(2-c))$. The initial value $h_0=1$
(the unstable fixed point), corresponds to the choice
$\beta=\beta_c={{8(2-c)}/{c^2}}$. In summary
\begin{equation}
h_{n} = (\beta/\beta_{c}) \xi^{n} \chi_{n} \ .
\label{eq:htochi}
\end{equation}
and 
the recursion formula then becomes a simple quadratic map (called
the ``$h$-map'' hereafter)
\begin{equation}
h_{n+1} = \xi h_{n} + (1-\xi)h_{n}^{2} \ ,
\label{eq:hmap}
\end{equation}
together with the initial condition $h_{0}= \beta/\beta_{c}$.
The restriction to $0<\beta<\beta_c$ corresponds to the
range $0<h_0<1$ which implies that for positive and finite $n$, $h_n$
stays within this interval. Note that Eq. (\ref{eq:hmap}) can be 
used to give $h_n$ as a function of $h_{n+1}$. This quadratic equation 
has two solutions; 
however, if we require $0<h_n,h_{n+1}<1$, only one solution
is acceptable and a unique inverse can be obtained by this restriction.
If we impose this restriction, the term ``group'' in RG can be understood 
in its proper sense.

We now discuss the
two fixed points.
The fixed point $h_0=0$ corresponds to 
the choice $\beta=0$ and is called the 
high-temperature (HT) fixed point.  
Remembering that $0<\xi<1$,
we see that the HT fixed point 
is stable, with eigenvalue $\xi$ in the linear approximation.
Using a graphical representation of the quadratic map, one sees that
the HT fixed point is globally attractive for the interval (0,1).
At the other end, the fixed point 
$h_0=1$ corresponds to the choice $\beta=\beta_c$, and is called the 
critical point. This fixed point  
is unstable, with eigenvalue $\lambda= (2-\xi)$.
Note that if $\xi$ is fixed by our initial choice of the 
dimensionality parameter $D$, the value of $\lambda$ can be seen
as an approximate value for largest
eigenvalue $\lambda_1$ of the hierarchical model.
This value is not too far off numerically. For instance for  $D=3$,
$2-\xi\simeq 1.37$ which can be compared with the 
known\cite{hier,rapid} value $1.42717\ldots$ .

We can expand the $h$-map about the unstable fixed point by 
using the reparametrization $h_{n} = 1-d_n$.
Note that
$d_{0} = (\beta_c-\beta)/\beta_{c}$ is the variable which appears in
the parametrization of the susceptibility given by Eq. (\ref{eq:param}).
The recursion formula for $d_n$ reads:
\begin{equation}
d_{n+1} = \lambda d_{n} + (1-\lambda) d_{n}^{2} \ ,
\label{eq:rec}\end{equation}
with $\lambda=2-\xi$.
We will call this map the ``$d$-map''. 
This map can be seen as a one variable version of Eq. (\ref{eq:quad}).
Note the similarity with the
original $h$-map. One can introduce a duality relation 
between the two maps which interchanges
$h_{n} \leftrightarrow d_{n}$ and  $\xi  \leftrightarrow \lambda$.
In the following, we use the notations $h_n=\tilde{d_n}$ and 
$\xi=\tilde{\lambda}$ to express the quantities appearing in the 
$h$-map as dual to the one appearing in the $d$-map.
If the duality transformation is applied twice, one returns to
the original quantities. 

Recalling that $0<h_0<1$, we also have $0<d_0<1$ with small values 
(approaching 0 from above)
in one variable corresponding to ``large'' values  (approaching 1 
from below) values  in the dual variable.
We would like to construct an expression for 
$\chi$ which is accurate for both small and large values of $d_0$.
In order to do this, we need
to use Eq. (\ref{eq:rec})
beyond the linear approximation. In the linear approximation, which
is justified when $d_{0}$ is very small ($\beta$ close to
$\beta_{c}$), $d_{n} \simeq \lambda^{n} d_{0}$. The linear approximation 
breaks down for values of $n$ of order 
$n^{\ast}$ defined by the relation $\lambda^{n^{\ast}} d_{0} = 1$. 
For $n$ larger than $n^{\ast}$, the non-linear terms become important
and $d_n$ approaches 1 from below 
as dictated by the global attractiveness of the HT fixed point. 
For $n$ large enough, the linearized $h-$map can be used to show 
that the HT fixed point is reached exponentially fast.
The linearization about the critical point 
provides the usual type of expression for 
the critical exponent $\gamma$: since  $\chi \sim \xi^{-n^{\ast}}$,
\begin{equation}
\gamma = -\ln \xi/\ln \lambda \ .
\label{eq:gamma}
\end{equation}

In order to refine the order of magnitude
estimate given by the leading singularity, we will
express $d_n$ as a function of $d_0$.
For this purpose, we
first construct a function $y(d) $ which transforms covariantly under 
Eq. (\ref{eq:rec}) :
\begin{equation}
y(\lambda d + (1-\lambda) d^{2})=\lambda y(d)\ .
\end{equation}
If we add the requirement that for small values $d$, 
$y(d)\simeq d$, this equation has a unique solution 
as a power series in $d$:
\begin{equation}
y(d) = d + {d^{2}\over\lambda} + \frac{2d^3}{\lambda(\lambda +1)}
+{{1+5\lambda^2}\over{\lambda^2(1+\lambda)(1+\lambda+\lambda^2)}}+
\ldots\ 
\label{eq:yofd}
\end{equation}
The inverse function can be constructed similarly.
In both cases, the coefficients can be calculated by simple recursion
relations, easily implementable on a computer.
We can now write:
\begin{equation}
d_{n}=y^{-1}(\lambda^ny(d_0)) \ .
\label{eq:dsubn}
\end{equation}

The idea of using intermediate variables with simple transformation
properties has a long history, for instance the angle-action variables in 
hamiltonian mechanics and the normal form of differential equations
appearing in Poincar\'e's dissertation.
For continuous RG transformations, Wegner\cite{wegner} introduced
the notion of scaling variables. In the case discussed here, the
construction of $y$ can be seen as a discrete version of 
Wegner's procedure. 

Much can be said about
the convergence properties of $y(d)$ and its inverse.
A numerical analysis of the coefficients indicates very clearly 
that $y^{-1}$ is an entire function,
while $y$ is analytical on the open disk of radius 1 
and has a power singularity when 
its argument tends to 1. Consequently, when  $0<d_0<1$, one can always
find an accurate expression for $d_n$ by using sufficiently many terms
in the expansions of $y$ and $y^{-1}$. Note that Eq. (\ref{eq:dsubn})
can also be used at negative values of $n$, providing the inverse  
transformations, which can be uniquely defined by requiring -- as in the 
case of the $h$-map discussed above -- that the preimage lies in the 
(0,1) interval. With this requirement, 
the transformation of Eq. (\ref{eq:rec})
becomes a group and the function $y(d)$ a non-unitary representation 
of this group. Note that one could also define a continuous transformation
by extending $n$ to all the real values.

Everything we have done for the $d$-map can be repeated almost verbatim 
for the $h$-map. We can construct a dual function $\tilde{y}(\tilde{d})$ 
transforming covariantly under the $h$-map with an expansion in $\tilde{d}$
of the form of the one given in Eq. (\ref{eq:yofd}) but with $\lambda $ 
replaced by $\tilde{\lambda}=\xi$. As for $y(d)$, 
we have clear numerical indication that $\tilde{y}(\tilde{d})$
is analytical on the open disk of radius 1 
and has a power singularity when its argument tends to 1. On the other
hand, its inverse is not entire but has a finite radius  
of convergence
with a square root behavior at the intersection of the 
boundary of the disk of convergence and the negative real axis.
Recalling that $0<\xi<1$ and that $\tilde{d_0}=h_0=\beta/
\beta_c$ we see that for $n$ large enough, we can use the above-described
expansions to calculate
\begin{equation}
h_{n}=\tilde{y}^{-1}(\xi^n\tilde{y}(h_0))\ .
\end{equation}
In the limit  where $n$ becomes infinite, the argument of  $\tilde{y}^{-1}$
goes to zero and it is justifiable to retain only 
the first term of its expansion.
Using the definition of the susceptibility of Eq. (\ref{eq:htochi}), we find 
that the $\xi$-dependence cancels and that we obtain the 
high-temperature expansion: 
\begin{equation}
\chi\equiv lim_{n->\infty}\chi_n={{\tilde{y}(h_0)}\over {h_0}}=
1+\beta{c\over{4(2-c)}}+ \ldots
\end{equation}
This expansion has features which are in qualitative agreement
with the actual HT series\cite{osc,high} of the hierarchical model. 

One can in principle use this HT expansion to extract the leading 
and subleading singularities of $\chi$.
However, this procedure is in general very inefficient 
because small physical effects
can be amplified dramatically in this expansion. 
For instance for $\lambda=1.8$, the sequence of ratios of successive
coefficients is completely ``noisy'' and no information can be extracted 
from it.
When $\lambda$ is lowered, the ``noise'' decreases and takes the form of
smooth log-periodic oscillating terms as in the examples 
discussed in Ref. \cite{osc}. 

Instead of using the HT expansion, we would like to have an expansion
in terms the dual variable $d_0$.
Such a goal can be achieved by combining the two covariant 
quantities $y$ and
$\tilde{y}$ into one invariant quantity which we call $A$ below.
Using the definition of $\gamma$ given in Eq. (\ref{eq:gamma}) , one sees
that $\lambda^{\gamma }=\tilde{\lambda}^{-1}$ and consequently
\begin{equation}
A=(y(d_n))^{\gamma} \tilde{y}(h_n)
\label{eq:adef}
\end{equation}
is $n$-independent. $A$ can be called 
a constant of motion or an RG invariant. We can now rewrite
\begin{equation}
\chi={A\over {(1-d_0)(y(d_0))^{\gamma}}} \ .
\label{eq:chido}
\end{equation}
We will show later that $A$ is a bounded function for $0<d_0<1$.
Eq. (\ref{eq:chido}) makes us suspect
that $\chi$ has a singularity when $d_0$ becomes close to 1, or
in other words, when $\beta$ becomes small. On the other hand, we know
that in this limit $\chi=1$. 
This apparent difficulty 
can be resolved by noticing that $\tilde{y}$ has a singularity
with power $-\gamma={\ln\tilde{\lambda}/{\ln \lambda}}$, consequently 
the dual quantity $y$
has a power singularity with dual exponent: $\tilde{\gamma}=1/\gamma$. 
Consequently, $(y(d_0))^{\gamma}$ at the denominator 
cancels the singularity and the expansion extends globally.

We now calculate
$A$ expressed as a function of $y(d_0)\equiv y_0$.
The invariance of $A$ under a RG transformation implies the discrete
scale invariance:
\begin{equation}
A(\lambda y_{0}) = A(y_{0}) \ ,
\label{eq:scaleinv}
\end{equation}
and the Fourier mode expansion:
\begin{equation}
A(y_0)=\sum_{n=-\infty}^{+\infty}a_ny_0^{in\omega}\ ,
\end{equation}
with $\omega=2\pi/\ln \lambda$,
and consequently
\begin{equation}
a_n={1\over{ \ln \lambda}}\int_{y_{a}}^{\lambda y_{a}}dy_0 
y_0^{-1-in\omega}A(y_0)\ .
\end{equation}
The lower value $y_{a}$ of the integration interval is arbitrary and we 
we can choose it
at our convenience and construct a decent series approximation 
for $A$ which is accurate in the integration interval. 
More explicitly, we can rewrite 
\begin{equation}
A(y_0)=(y_0)^{\gamma} \tilde{y}(1-y^{-1}(y_0)))\ ,
\label{eq:aexp}
\end{equation}
and use the series expansions for $y^{-1}$ and $\tilde{y}$.
If $y_0$ is small, we need a few terms for $y^{-1}$ and many for 
$\tilde{y}$.
If $y_0$ is large, we need many terms for
$y^{-1}$ and a few terms for $\tilde{y}$. 
These two extreme possibilities are very inefficient ways to calculate 
the Fourier coefficients. We have compared 
the approximate values $a_0(m,\tilde{m})$ 
obtained from expansions of Eq. 
(\ref{eq:aexp}) with $m$ terms for $y^{-1}$ and $\tilde{m}$ for $\tilde{y}$
with an accurate value of $a_0$ and found the approximate
behavior
\begin{equation}
|a_0(m,\tilde{m})-a_0|\propto {\rm exp}(-K_1(m+\tilde{m})+K_2(m-\tilde{m})^2)
\ ,
\label{eq:optim}
\end{equation}
where $K_1$ and $K_2$ are positive constants. 
Eq. (\ref{eq:optim}) implies that for $m+\tilde{m}$ fixed, it is 
very advantageous to pick the ``self-dual'' option $m\simeq\tilde{m}$.

\begin{figure}
\centerline{\psfig{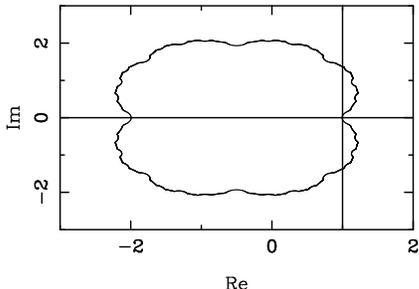}}
\caption{Boundary of the basin of attraction of 0 
for the complex $h$-map with $\xi$=0.5. }
\label{fig:julia}
\end{figure}

For values of $\xi$ not too small, the contribution
of 
the non-zero Fourier modes to the susceptibility is
exponentially suppressed. We found indications for the following behavior:
\begin{equation}
|a_n|/|a_0|\propto {\rm exp}(-|n|\pi \omega /2)\ ,
\label{eq:expsup}
\end{equation}
as in another 
example of function with log-periodic 
oscillations discussed in section 5 of Ref. \cite{osc}.
The first indication is the shape of the basin of attraction 
of the stable fixed 
point of the $h$-map, 
in the complex $h$-plane. Near the unstable fixed point 
($h=1$, $d=0$), we can linearize 
$y\simeq d=1-h$. From Figure \ref{fig:julia}, 
we see that if $d=-\delta{\rm exp}(i\theta)$ with $0<\delta<<1$, the 
points such that $|\theta|<\pi/2$ are attracted to zero. Given
the behavior of $y^{in\omega}$ in this region, this requires for large
values of $n$
the suppression given by Eq. (\ref{eq:expsup}).
We have also checked explicitly for $n=1$ and 2, that this 
exponential suppression provides a good fit of the data
for $\omega >11$.
Despite the suppression, 
the non-zero modes are quite visible in the HT expansion
because of a factor $(\Gamma(\gamma+in\omega))^{-1}$ which appears in the 
the expression of the HT coefficients (see Eq. (3.7) of Ref.\cite{osc})
and cancels, in leading order, the suppression from Eq. (\ref{eq:expsup}).

The construction of covariant
quantities extends easily to the case of several variables
and can be used to attempt accurate calculations with 
the hierarchical model. The 
extension to the case of nearest neighbor model is a more 
difficult procedure. It is clear that one can as in Ref.\cite{finite} 
use polynomial
approximations for the Fourier transform of the local measure
and write a recursion formula for a block-spin transformation. However
this procedure generates non-local interactions. It is not clear
if these interactions can be exponentiated in a compact way 
(as in the Gaussian case). The present work should be seen as
an encouragement to attack
this difficult question.

Two remarks can be made regarding the construction of covariant
quantities in the multivariable case.
The first one is that for a $l-$dimensional quadratic
recursion formula, the number of coefficients to be determined at
order $m$ grows like $l^m$ and consequently optimization is an 
important consideration. Second, the problem has no solution 
if a given eigenvalue can be written exactly as a product of other eigenvalues.
This is the exponentiated form of the problem of logarithmic anomalies
raised by Wegner in Ref.\cite{wegner}. Generically, such a problem
is likely to occur is an approximate way. For instance, if we use the 
eigenvalues for the $D=3$ hierarchical model given in Ref. \cite{rapid},
we find that $\lambda_3-\lambda_2^5\simeq 10^{-2}$. When this is the case, 
we have a ``small denominator problem'' which reflects approximate
``resonance'' among the various ``modes'' present. 
Pursuing this analogy, the results presented here provide
a solution of a non-linear problem with one degree of freedom.
Their application to realistic systems seems likely to have a
complexity and an interest comparable to systems of coupled 
non-linear oscillators.  

We thank the participants to the 
Math-Physics seminar at the U. of Iowa for extended discussions.
This research was supported in part by the Department of Energy
under Contract No. FG02-91ER40664.

\end{multicols}

\end{document}